\documentclass[12pt]{article}
\begin{document}
\noindent
\textbf{Manifestly covariant current matrix elements in the Point 
 Form Relativistic Hamiltonian Dynamics }

\vskip 1.0 truecm

\centerline{M. De Sanctis $^{a,~b}$}
\vskip 1.0 truecm
\noindent
\textit{$^a$ Departamento de F\'isica, Universidad Nacional de Colombia,
Bogot\'a D. C., Colombia.  }

\noindent
\textit{$^b$ INFN sez. di Roma, P.le A. Moro 2, 00185 Roma, Italy. }

\noindent
e-mail :  mdesanctis@unal.edu.co  and   maurizio.desanctis@roma1.infn.it

\vskip 1.0 truecm
\begin{abstract}
\noindent
A manifestly covariant expression for the current matrix elements of three quark
bound systems is derived in the framework of the 
Point Form Relativistic Hamiltonian Dynamics.
The relativistic impulse approximation is assumed in the model.
A critical comparison is made with other expressions usually given in the literature.

\vskip 0.25 truecm
\noindent
PACS number(s): 11.30.Cp, 24.10.Jv, 13.40.Gp
\end{abstract}

\vskip 0.5 truecm
\noindent
\textbf{1. Introduction}
\vskip 0.5 truecm
\noindent
Aim of the present work is to show that, in the context of the Point Form Relativistic
Hamiltonian Dynamics (PF RHD), it is possible to write, for hadronic bound systems,
\textit{ manifestly covariant} 
matrix elements of the current operators. 

\noindent
For clarity, we \textit{define} as 
\textit{ manifest covariance} the property of an equation of being written in terms of
quantities that (i) transform in a well-known way under Lorentz transformations and
(ii) are not related to a specific reference frame.

\vskip 0.5 truecm
\noindent
We  use  a \textit{relativistic impulse approximation} [1,2] (RIA), 
that generalizes at a relativistic level
the widely known model used
for the  study of the electromagnetic interactions of nuclear systems.

\noindent
We shall examine in more detail the case of the
four-vector  electromagnetic current, but the method has been developed to
study any kind of interaction, in particular  the axial-vector case [3,4] that is 
relevant for  the weak structure of the hadrons.

\vskip 0.5 truecm
\noindent
In this article we shall specifically refer to the nucleon as a system composed 
by $N=3$ (identical), $s=1/2$, constituent quarks. 
The  same procedure can be immediately generalized  to the study of the $N\neq 3$
composite systems.
The cases of   nonidentical constituents and  $s\neq 1/2$ 
will be studied, within the same theoretical framework, in subsequent works.

\vskip 0.5 truecm
\noindent
We also show that our model, 
that is the result of an independent investigation [5],
is completely equivalent to the standard formalism (SF) 
of PF RHD, developed in refs.[1-3] and implemented with great clarity
and precision in ref.[6].
In this concern we anticipate that the main differences between our model and  SF
are the following: 

\noindent
(a) we use, in our work, the formalism of the Dirac equation boosting
covariantly the Dirac spinors,
while in SF Wigner rotation matrices (not manifestly covariant) are employed;

\noindent
(b) the spatial part of the bound system  four-velocity 
and the (independent) three-momenta of two quarks
are used in our work as spatial variables for the representation (or projection) states, 
while the \textit{velocity states}
are considered in SF; the use of these states, that are related to the rest frame
of the bound system does not fulfill the requirement (ii) given above to have
manifestly covariant equations;

\noindent
(c) furthermore, we perform \textit{manifestly covariant} integrations over that
spatial variables to calculate the matrix elements of the current operators.
\vskip 0.5 truecm
\noindent
We highlight that the electroweak matrix elements calculated by means of PF RHD
allow to reproduce with good accuracy the experimental nucleon form factors [2-7]. 

\vskip 0.5 truecm
\noindent
The model studied in the present work allows for two further developments
(to be studied in different works): 
the definition of a dynamically conserved electromagnetic current by means of
a suitable minimal coupling procedure and the introduction of (virtual) negative
energy states in the current matrix elements.

\vskip 0.5 truecm
\noindent
The paper is organized as follows.

\noindent
In sect.2 we revise the construction of the Poincar\'e algebra generators, introducing, 
at the same time, the operators that  are used to describe the dynamics of 
the bound system.

\noindent
In sect.3, by introducing the projection states of our model,
we define  the wave functions for the bound system in the framework 
of PF RHD, also discussing their boost properties.

\noindent
In sect.4 we explicitly construct the Dirac wave functions of our formalism, 
showing the equivalence of their boost properties with those of the SF.

\noindent
Finally, in sect.5, the matrix elements of the current operators are studied
by means of the RIA. The main result of this work is our covariant expression 
given in eq.(5.3c).
An accurate comparison with SF is performed transforming eq.(5.3c) into 
the standard form of eq.(5.20).

\vskip 0.5 truecm
\noindent
\textbf{2. The Poincar\'e Algebra}
\vskip 0.5 truecm
\noindent
In the present work, considering  particles of mass $m$,
\textit{always on-shell}, 
we transform the  four-momentum
$p^\mu=(\epsilon({\bf p}),{\bf p})$ by means of a
\textit{ canonical boost} written in the following standard form

$$ \epsilon_b({\bf p};{\bf v})=
 \epsilon({\bf p}_b({\bf p}; {\bf v}))= v^0 \epsilon({\bf p})+{\bf v p} 
\eqno{(2.1a)  }$$
$$ {\bf p}_b({\bf p};{\bf v})= 
{\bf p}  +  {\bf v}( \ {\bf v} {\bf p}
\frac{1}{v^0+1} +\epsilon({\bf p})) \eqno{(2.1b)}$$
The two previous equations are usually resumed in the form
$$ p^\mu_b=L^\mu_{~\nu}({\bf v})p^\nu  \eqno{ (2.1c)}$$

\vskip 0.5 truecm
\noindent
In eqs.(2.1a,b) we have introduced the time component of the four-momentum 
of the particle, i.e. the energy, 
as  $\epsilon({\bf p})=[{\bf p}^2 + m^2]^{1/2}$ 
and $v^\mu=([{\bf v}^2 +1]^{1/2}, {\bf v})$, 
that is the four-velocity boost parameter. We recall that the \textit{ physical}
velocity of the initial frame measured from the boosted one is
${\bf u}={\bf v}/ {v^0} $. 

\noindent
The \textit{independent} transformation equation, that is
used to define the boost in the Hilbert space, is eq.(2.1b), while eq.(2.1a) 
can be obtained from that one by calculating the on-shell energy of the particle with 
the boosted momentum.

\vskip 0.5 truecm
\noindent
As anticipated, we follow the scheeme of the PF RHD when defining the generators
of the Poincar\'e algebra [1]. In more detail,
for a system of three quarks, the total angular momentum $\bf J$ 
and the total boost $\bf K$, being \textit{free} of the interaction,
are written as the \textit{sum} of the single particle generators, in the form:

$${\bf J}=\sum_{i=1}^3 \left({\bf r}_i \times {\bf p}_i + {\bf s}_i
\right) \eqno{(2.2a)}$$

$${\bf K}=\sum_{i=1}^3 \left[  
 \frac{1} {2} ({\bf r}_i\epsilon({\bf p}_i)+\epsilon({\bf p}_i){\bf r}_i)
 + \frac{{\bf p}_i \times {\bf s}_i} {\epsilon({\bf p}_i) +m} \right] \eqno{(2.2b)} $$
where ${\bf p}_i$, ${\bf r}_i$, ${\bf s}_i= {\frac {1} {2}} \vec \sigma_i$,
$m$ and  $\epsilon({\bf p}_i)$ 
respectively represent the three-momentum, the conjugated (position) variable,
the spin, the mass and the energy of the i-th quark.

\noindent
For completeness we also give the expression of the finite boost operator, that in the
PF RHD  is not modified by the interaction:

$$B({\bf v})=\exp(i{\bf K \cdot U}) \simeq 1+i\delta {\bf u}\cdot {\bf K} \eqno{(2.2c)}$$
with
$${\bf U} =\frac {{\bf v}} {|{\bf v}|} \tanh^{-1}( {|{\bf v}|\over v^0} ) =
 \frac {{\bf u}} {|{\bf u}|} \tanh^{-1}( {|{\bf u}|} ) \eqno {(2.2d)}$$

\vskip 0.5 truecm
\noindent
On the other hand, the total four-momentum  operator of the system, that is
$P^{\mu}=(P^0=H, {\bf P})$ depends on the interaction among the constituent quarks.

\noindent
We shall  define the operator  $P^{\mu}$ in eq.(2.19).
To this aim, we have, previously,  to introduce:

\noindent
(a) the quantum-mechanical operator $V^\mu$,
that represents the four-velocity of the bound system measured from 
a generic reference frame (GF); 

\noindent
(b) the other dynamical variables of the 
quantum mechanical model. 
\vskip 0.5 truecm
\noindent
We first consider point (a), that is the construction 
of the \textit{ quantum mechanical operator} $V^\mu$.
In order to help the reader to understand the physical meaning of the following
procedure,
we note that the four-momentum $P^\mu$
of a system, as a \textit{classical quantity}, can be written in terms of $V^\mu$ 
as
$$P^\mu=M V^\mu \eqno(2.3)$$ 
where the physical mass $M$ of the (bound) system has been introduced.
The corresponding quantum-mechanical expression will be given in eq.(2.19).

\noindent
To derive this expression,  we have to write $V^\mu$ as a function
of the momenta of the constituents.
As first step, we introduce the  rest frame (RF) 
four-momentum of the i-th quark
$$ p_i^{* \mu}=(\epsilon({\bf p}_i^*), {\bf p}_i^*) \eqno(2.4)$$
Here and in the following, the asterisk denotes the quantities observed in the RF.
The sum of the four-momenta of the three constituent quarks,  
is, by definition of the RF 

$${\sum_{i=1}^{3}}  p_i^{* \mu}=
( {\sum_{i=1}^{3}} \epsilon({\bf p}_i^*)=M_f, {\bf 0}) \eqno(2.5)$$
where we have also introduced $M_f$ that represents the \textit{free mass operator} 
of the system.
By applying the Lorentz transformation of eqs.(2.1a,b) 
(as a function of the parameter $\bf V$) 
to the $ p_i^{* \mu}$ and
also using eq.(2.5), one can write the sum 
of the four-momenta of the particles
in a GF as

$$\sum_{i=1}^3 p_i^{ \mu}= V^{\mu}M_f \eqno(2.6) $$
with
$$ p_i^\mu=(\epsilon({\bf p}_i), {\bf p_i})$$
\noindent
We highlight that $M_f$, as defined in eq.(2.5),
is a \textit{nonvanishing and Lorentz invariant} quantity.
The nonvanishing character of $M_f$ allows to solve the previous 
equation with respect to $V^{\mu}$. 
Lorentz invariance allows to write $M_f$
in terms of the $p_i^{ \mu}$ observed in a GF.
In this way one can express $V^{\mu}$ as
a function the $p_i^{ \mu}$, or, more precisely, of the three-momenta
${\bf p}_i$.

\noindent
(Note that, the ${\bf p}_i$, with $i=1,2,3$ represent, in the \textit{first step} 
of the construction, the spatial dynamical variables
of the relativistic model.
The \textit{final choice} of the spatial variables will be given in the following.)

\noindent
In more detail, $M_f$ is expressed as a function 
of the momenta in a GF in the form

$$M_f=M_f({\bf p}_1,{\bf p}_2, {\bf p}_3     )=
\left[ 
\sum_{i j=1}^3 p_i^{ \mu} 
p_j^{ \nu} g_{\mu \nu}
        \right]^{1/2 } \eqno{(2.7)}$$
that will be taken as the definition of the \textit{operator} $M_f$.
In consequence,we can also write

$$V^\mu( {\bf p}_1,{\bf p}_2, {\bf p}_3 )=
 [M_f({\bf p}_1,{\bf p}_2, {\bf p}_3 ) ]^{-1}
\sum_{i=1}^3  p_i^{ \mu}
\eqno(2.8a)$$
and , obviously
$$V^\mu V_\mu=1 \eqno(2.8b)$$
$$V^0({\bf V})=[1+{\bf V}^2]^{1/2} \eqno(2.8c)$$
Let us note that the observable four-vector 
$ V^\mu$, as given in eq.(2.8a),
transforms in the same way as a standard four-momentum, that is
replacing $\bf p$ with $\bf V$, $\epsilon({\bf p})$ with $V^0({\bf V})$
in eq.(2.1b). In this way we introduce
$$V^0_b=V^0_b({\bf V};{\bf v})=V^0({\bf V}_b({\bf V} ;{ \bf v}))\eqno(2.9a)$$
$${\bf V}_b= {\bf V}_b({\bf V} ;{\bf v})\eqno(2.9b)$$ 
This result, that is also consistent with 
eq.(2.3), can be easily derived by transforming,
with the help of eqs.(2.1a,b),
the $ p_i^{ \mu} $ that appear in eq.(2.8a).
\vskip 0.5 truecm
\noindent
As for point (b),
we  can now introduce the \textit{final choice} for the complete set 
of commuting \textit{operators} that will be used
for the quantum mechanical description of the system. 
To this aim we note that,
due to its definition in eq.(2.8a), 
the operator $V^\mu$ commutes
with the momenta of all the particles. In consequence, 
it is possible to choose the following operators:

\noindent
(i) as spatial variables,
the three-momenta of $2$ quarks, say 
${\bf p}_2, {\bf p}_3 $,
and the spatial components of the four-velocity $\bf V$; those variables replace
the \textit{first step} choice of $ {\bf p}_1,{\bf p}_2, {\bf p}_3 $;     

\noindent 
(ii) the spin operators of the three quarks; the eigenvalues of their projections 
on the z axis will be denoted as
$\sigma_1,\sigma_2,\sigma_3 $.

\vskip 0.5 truecm
\noindent
For further developments, it is necessary to express $p_1^\mu$ and $M_f$ as functions of
${\bf p}_2, {\bf p}_3 $ and ${\bf V}$.

\noindent
First, we recall that the rest frame quark energies are \textit{invariant}
quantities [8], that can be written as
$$\epsilon_i^*=\epsilon({\bf p}_i^*)=V_\mu p_i^{ \mu}\eqno(2.10)$$

\noindent
Second, we write eq.(2.6) in the form

$$ p_2^\mu + p_3^\mu = -p_1^\mu + {V^\mu}
\cdot[ \epsilon({\bf p}_1^*)+\epsilon({\bf p}_2^*)+\epsilon({\bf p}_3^*)]\eqno{(2.11)}$$
Then, squaring both sides, with the help of eq.(2.10), one obtains the RF energy 
of the quark $\# 1$ as a function of ${\bf p_1}$, ${\bf p_2}$ and ${\bf V}$:
$$\epsilon({\bf p}_1^*)=\epsilon_1^*({\bf p}_2, {\bf p}_3, {\bf V})=$$ 
$$\left[m^2 -(p_2^\mu+p_3^\mu)(p_2^\nu+p_3^\nu)g_{\mu\nu}
+[(p_2^\mu + p_3^\mu)V_\mu]^2 \right]^{1/2} \eqno(2.12) $$
where $p_2^\mu$, $p_3^\mu$ and $V^\mu$ are functions of 
${\bf p}_2$, ${\bf p}_3$, and ${\bf V}$, respectively.

\noindent
Finally, we find
$$M_f({\bf p}_2,{\bf p}_3,{\bf V})= (p_2^\mu + p_3^\mu)V_\mu 
+\epsilon_1^*({\bf p}_2, {\bf p}_3, {\bf V}) \eqno(2.13)$$ 
and, by means of eq.(6)
$$p_1^\mu({\bf p}_2,{\bf p}_3,{\bf V})= - (p_2^\mu + p_3^\mu)
+ V^\mu \cdot M_f({\bf p}_2,{\bf p}_3,{\bf V}) \eqno(2.14)$$
By definition, $M_f$ is a Lorentz invariant operator, that is
$$[{\bf K}, M_f]=0\eqno(2.15)$$
We now introduce the interaction among the quarks by means of
the total mass operator $M$ that, according to the Bakamjian-Thomas construction [1,9],
is defined as
$$ M= M_f+ W \eqno(2.16)$$
where $W$ reprents a \textit{Lorentz invariant} interaction operator, that means
$$[{\bf K}, W]=0\eqno(2.17a)$$
and, in consequence,
$$[{\bf K}, M]=0\eqno(2.17b)$$

\noindent
In this work, we do not enter into the details of the definition of $W$. We only
point out that  rotationally scalar operators, defined in the RF 
(as the phenomenological potentials generally used for the relativized constituent 
quark models, in particular the hypercentral potentials [10]),
are formally Lorentz invariant and can be also written in an explicit
invariant form by means of the dynamical variables of the model.

\noindent
Note that, if the interaction operator $W$ represents a quasi-potential
derived from an underlying field theory, its expression is, in general,
highly momentum dependent.

\noindent
In any case, being defined in the RF, the interaction operator $W$ has nonvanishing
matrix elements only between states with the same $V^\mu$,
that means
$$[V^\mu, W]=0\eqno(2.18)$$

\vskip 0.5 truecm
\noindent
We can now introduce the generators of the time and space 
translation of the system, that is the four-momentum operator, as
$$P^\mu= M\cdot V^{\mu}=(M_f+W)\cdot V^{\mu}\eqno(2.19)$$
that is the same expression of eq.(2.3), but considered as a
definition of a quantum mechanical operator.

\vskip 0.5 truecm
\noindent
Standard calculations [1] show that the total generators defined in eqs.(2.2a,b)
and (2.19) fulfill the Poincar\'e group commutation rules [11,12].
\vskip 1.0 truecm
\noindent
\textbf{3. The wave functions of the model}
\vskip 0.5 truecm
\noindent
We  now turn to introduce the
representation states that will be used to write down explicitly the wave functions
of the model. 
Following the definitions of the dynamical variables given in the previous section, 
one has

$$ |\psi_r>= |{\bf p}_2,{\bf p}_3, {\bf V}; \sigma_1,\sigma_2,\sigma_3> \eqno(3.1) $$
with the normalization 
$$<\psi_r|\psi_r'>=
<{\bf p}_2,{\bf p}_3, {\bf V}; \sigma_1,\sigma_2,\sigma_3|
{\bf p}_2',{\bf p}_3', {\bf V}'; \sigma_1',\sigma_2',\sigma_3'>=$$
$$\delta({\bf p}_2'-{\bf p}_2)\delta({\bf p}_3'-{\bf p}_3)\delta({\bf V}'-{\bf V})
\delta_{{\sigma_1}' \sigma_1} \delta_{{\sigma_2}' \sigma_2} \delta_{{\sigma_3}' \sigma_3}
\eqno(3.2)$$ 
The choice of eq.(3.1), as it will be shown in sect.4,
helps to introduce in a very clear way 
the relativistic impulse approximation
for the current matrix elements. 
On the other hand, in SF 
a different type of representation states, currently denoted as
\textit{velocity states}, is generally used to study 
the relativistic bound state wave functions. In the velocity states
the spatial variables are represented by $\bf V $ and by the three (not
indepedendent) rest frame momenta 
${\bf p}_1^*,   {\bf p}_2^*,    {\bf p}_3^*     $ 
or better by the two
(independent) Jacobi momenta ${\bf p}_{\rho}, {\bf p}_{\lambda} $. 
As shown in ref.[2],
the Lorentz transformation  of these states is given by the standard boost
of $\bf V$ and by a Wigner rotation of the 
rest frame momenta or of the Jacobi momenta.
If  also the spin projections are referred to the RF, the same Wigner rotation acts
on the spin variables.
\vskip 0.5 truecm
\noindent
By applying the boost operator of eq.(2.2c) to the representation states, 
one obtains:
$$B({\bf v}) |{\bf p}_2,{\bf p}_3, {\bf V}; \sigma_1,\sigma_2,\sigma_3> =$$
$$R({\bf p}_1; {\bf v})R({\bf p}_2; {\bf v})R({\bf p}_3; {\bf v})
|{\bf p}_{2b},{\bf p}_{3b}, {\bf V}_b; \sigma_1,\sigma_2,\sigma_3> 
G({\bf p}_2,{\bf p}_3, {\bf V};{\bf v})
\eqno (3.3a) $$
with
$$G({\bf p}_2,{\bf p}_3, {\bf V};{\bf v})=[\frac {\epsilon_b ({\bf p}_2; {\bf v})}
{\epsilon ({\bf p}_2)}
 \frac {\epsilon_b ({\bf p}_3; {\bf v})}
{\epsilon ({\bf p}_3)}
 \frac {V^0_b ({\bf V}; {\bf v})}
{{\bf V}^0 ({\bf V})}
 ]^{1/2} \eqno(3.3b)$$

\noindent
The previous equations show that the action of the boost operator 
on the representation states can be divided into a spatial (a) and a spin (b) part.
\vskip 0.5 truecm
\noindent
(a) The spatial part, denoted in the following as $\widehat B({\bf v})$, 
produces an eigenstate of the boosted momenta 
$ {\bf p}_{2b},{\bf p}_{3b}, {\bf V}_b$ that are  taken as functions 
of the corresponding unboosted variables by means of eq.(2.1b); 
the numerical  factor  $G({\bf p}_2,{\bf p}_3, {\bf V})$
is due to the nonlinearity, with respect to the momenta, 
of the boost generator of eq.(2.2b) and provides for the correct normalization of
the  state, being  $\widehat B({\bf v})$ a unitary operator. To simplify further developments
we introduce the following spatial matrix element

$$ <{\bf p}_2,{\bf p}_3, {\bf V}|\widehat B({\bf v})
|{{\bf p}_2}',{{\bf p}_3}', {\bf V}'> = G({\bf p}_2,{\bf p}_3, {\bf V};{\bf v})$$
$$\delta( {\bf p}_2 - {\bf p}_{2b}( {{\bf p}_2}'; {\bf v}))
  \delta( {\bf p}_3 - {\bf p}_{3b}( {{\bf p}_3}'; {\bf v}))
  \delta( {\bf V}   - {\bf V}_b(  {\bf V}';    {\bf v}))\eqno(3.4)$$
\noindent
and recall the following property of the delta functions

$$\delta( {\bf p}_i - {\bf p}_{ib}( {{\bf p}_i}'; {\bf v}))=
  \delta( {{\bf p}_i}' - {\bf p}_{ib}( {{\bf p}_i}; {-\bf v}))
{\frac  {\epsilon({{\bf p}_i}' )} {\epsilon_b ({{\bf p}_i}'; {\bf v})} }
\eqno(3.5)$$
with $i=2,3$. Note that   ${\bf p}_{ib}( {{\bf p}_i}; {-\bf v})$
represents the \textit{inverse} Lorentz transformation on ${{\bf p}_i}$
that is obtained using in eq.(1.b) the boost parameter $ {-\bf v}$.

\vskip 0.5 truecm
\noindent
(b) In eq.(3.3a) the spin part is given by the product of the 
$R({\bf p}_i;{\bf v})$ that represent
the Wigner spin rotation operators (due to the second term in the generator
of eq.(2.2b)) that depend on  the numerical values of the ${\bf p}_i$.
The (not independent) momentum ${\bf p}_1$ is obtained by means of eq.(2.14).
By considering the Pauli spinor representation for the spin states,
for further developments we introduce  the following  matrix elements

 $$ w_{{\sigma_i}'}^+  R({\bf p}_i;{\bf v})w_{\sigma_i}= 
R_{{\sigma_i}' \sigma_i}  ({\bf p}_i;{\bf v} ) \eqno(3.6)$$
\noindent
In SF the matrix elements of the spin rotation operators have been denoted as

$$ R_{{\sigma_i}' \sigma_i}  ({\bf p}_i;{\bf v} )=
 D_{{\sigma_i}' \sigma_i}^{1/2}[ R_W({\bf p}_i,B({\bf v}))] \eqno(3.7) $$ 
Such notation is used 
to represent the spin $1/2$ rotation matrices considered as functions of the
Wigner rotation related to the momentum ${\bf p}_i$ and to the boost $B({\bf v})$.

\vskip 0.5 truecm
\noindent
The wave function of our model is determined in the RF,
as a function of the Jacobi momenta ${\bf p}_\rho$, ${\bf p}_\lambda$
solving the mass eigenvalue
equation for the mass operator introduced in eq.(2.16). This solution is a
\textit{velocity state} solution with ${\bf V}={\bf 0}$, 
as indicated in the next equation by a Dirac $\delta$ function.
It is written as

$$\psi_{RF}(  {\bf p}_\rho, {\bf p}_\lambda,{\bf V})=
\psi^{J \Sigma}({\bf p}_\rho, {\bf p}_\lambda)\delta({\bf V} ) \eqno(3.8a)$$
with
$$\psi^{J \Sigma}({\bf p}_\rho, {\bf p}_\lambda)=
\sum_{\sigma_1 \sigma_2 \sigma_3} 
\psi^{J \Sigma }_{\sigma_1 \sigma_2 \sigma_3}  
({\bf p}_\rho, {\bf p}_\lambda)
w_{\sigma_1} w_{\sigma_2} w_{\sigma_3} \eqno(3.8b)$$
In the previous expression $J, \Sigma$ respectively represent 
the total angular momentum (absolute value) and its projection on the z axis.
This state is constructed by coupling the angular momenta with Clebsch-Gordan
coefficients,
for example  according to the standard scheeme [13]
$$[[l_\rho\otimes l_\lambda]^L \otimes S]^{J \Sigma} $$
with
$$[[s_1\otimes s_2]^{S_{12}}\otimes s_3]^{S M_S}$$
Note that, in  eq.(3.8b) the dependence on the quark Pauli
spinors has been highlighted in order to make a comparison with SF.

\noindent
For the following developments, it is convenient to introduce,
as spatial variables, instead of the Jacobi
momenta, the RF momenta ${\bf p}_2^*$, ${\bf p}_3^*$.
The former and the latter momenta are connected by a standard linear relation.
We have

$$\psi^{J \Sigma}({\bf p}_2^*, {\bf p}_3^*)= j^{1/2}
 \psi^{J \Sigma}
({\bf p}_\rho     ({\bf p}_2^*, {\bf p}_3^*), 
 {\bf p}_\lambda  ({\bf p}_2^*, {\bf p}_3^*)    )\eqno(3.9)$$
where $j^{1/2}$ represents the (numerical) constant factor 
that is used to keep the normalization to unity
for the wave function when using the new variables ${\bf p}_2^*, {\bf p}_3^*$.
The wave function of the previous equation can be decomposed 
with respect to the Pauli spinors in the same way 
as the wave function given in eq.(3.8b).
As before, 
$$\psi_{RF}({\bf p}_2^*, {\bf p}_3^*,{\bf V})= 
\psi^{J \Sigma}({\bf p}_2^*, {\bf p}_3^*) \delta({\bf V} ) \eqno(3.10)$$
We can now determine the wave function of the system in a GF boosting 
the RF wave function given in the previous equation. We use the boost parameter
${\bf v}_G$, that, as usual, represents the spatial part of the four-velocity of
the bound system observed from the GF. One has

$$\psi_G({\bf p}_2, {\bf p}_3, {\bf V})=< {\bf p}_2, {\bf p}_3, {\bf V}|
B({\bf v}_G)|\psi_{RF}> = $$
$$\int d^3{{\bf p}_2}' d^3{{\bf p}_3}' d^3{\bf V}'
< {\bf p}_2, {\bf p}_3, {\bf V}|B({\bf v}_G)|{{\bf p}_2}',{{\bf p}_3}', {\bf V}'>$$
$$<{{\bf p}_2}',{{\bf p}_3}', {\bf V}'|\psi_{RF}> \eqno(3.11)$$
By using the explicit expression of the RF wave function of eq.(3.10), the property
of the spatial part of the boost  and of the $\delta$ functions,
respectively given in eqs.(3.4) and (3.5) and, finally, the spin rotation operators
of eq.(3.6), one obtains 
$$\psi_G({\bf p}_2, {\bf p}_3, {\bf V})=
R({\bf p}^*_{1G}; {\bf v}_G)R({\bf p}^*_{2G}; {\bf v}_G)R({\bf p}^*_{3G}; {\bf v}_G)$$
$$ <{\bf p}_2, {\bf p}_3, {\bf V}|
\widehat B({\bf v}_G)|\psi_{RF}> 
 \eqno(3.12a)$$
where we have introduced the spatial part of the boosted wave function
$$ <{\bf p}_2, {\bf p}_3, {\bf V}|
\widehat B({\bf v}_G)|\psi_{RF}> =
\left[{ \frac  {\epsilon({{{\bf p}^*_{2G}}}) \epsilon({{{\bf p}^*_{3G}}}) }
               {\epsilon({\bf p}_2)   \epsilon({\bf p}_3)  }  }\right]^{1/2}
(1+{\bf v}_G^2)^{-1/4}$$
$$\psi^{J \Sigma}({\bf p}^*_{2G}, {\bf p}^*_{3G})
\delta({\bf V} - {\bf v}_G)   \eqno(3.12b)$$ 
also, by means eq.(2.1b), we have used
 $${\bf p}^*_{iG}= {\bf p}_b({\bf p}_i; -{\bf v}_G)\eqno(3.12c)$$
that represent the rest frame three-momenta considered as functions of 
the three-momenta of the GF, transformed by means of the parameter ${\bf v}_G$.
\vskip 1.0 truecm
\noindent
\textbf{4. The Dirac equation formalism}
\noindent
\vskip 0.5 truecm
\noindent
In order to construct operators that manifestly transform as Lorentz
tensors, it is very useful to make use  of the Dirac equation formalism.

\noindent
First,  we define the RF Dirac wave function in the form

$$\psi^D_{RF}({\bf p}_2^*, {\bf p}_3^*,{\bf V})= 
u({\bf p}_1^*) u({\bf p}_2^*) u({\bf p}_3^*)
\psi_{RF}({\bf p}_2^*, {\bf p}_3^*,{\bf V}) \eqno(4.1a)$$ 
with the \textit{Dirac spinors}, 
$$u({\bf p}_i) = {1\over\sqrt{2m}}
\left[ \matrix { \sqrt{\epsilon({\bf p}_i) + m}\cr &\cr
({\bf p}_i \vec{\sigma}_i ) \over{\sqrt{\epsilon({\bf p}_i) +m}}\cr }
\right]\eqno(4.1b)$$
For brevity, we  denote these quantities,
here and in the following,  as 
(positive energy) \textit{Dirac spinors}, 
taking into account that they represent $4\times 2$ 
\textit{ matrices} acting onto
the Pauli spinors $w_{{\sigma}_i}$ contained in 
$\psi_{RF}({\bf p}_2^*, {\bf p}_3^*,{\bf V})$. 
They are covariantly normalized as $\bar u({\bf p}_i) u({\bf p}_i) = {\bf 1}$.

\noindent
We recall that the Dirac spinors are boosted by means of the nonunitary Dirac
boost operator
$$B^D_i({\bf v})=[B^D_i({\bf v})]^+=
[{1\over 2}(v^0+1)]^{1/ 2}+
[{1\over 2}(v^0-1)]^{1/ 2} { ({{\bf v}{\gamma_i}^0 \vec\gamma_i}) \over {| {\bf v}|}}
  \eqno(4.2)$$
where we have introduced the Dirac the gamma matrices 
${\gamma_i}^\mu=({\gamma_i}^0,\vec\gamma_i)$
for the i-th particle; also, $v^0$ is the time component of the four-velocity
boost parameter. Standard calculations show the following very important
property of the Dirac boost when applied to the Dirac spinors

$$B^D_i({\bf v})u({\bf p}_i)= u({\bf p}_b({\bf p}_i; {\bf v}))
R({\bf p}_i; {\bf v}) \eqno(4.3)$$
It shows that the Dirac boost produces a Dirac spinor of the boosted momentum
applied to the spin rotation operator, given in eq.(3.7), that acts onto the 
Pauli spinor.
\vskip 0.5 truecm
\noindent
Introducing
$$B^D({\bf v}_G)=
B^D_1({\bf v}_G)\otimes B^D_2({\bf v}_G)\otimes B^D_3({\bf v}_G)\eqno(4.4)$$
we now construct the GF Dirac wave function 
for the three quark system
by means of the following boost

$$\psi^D_G({\bf p}_2, {\bf p}_3,{\bf V})= 
B^D({\bf v}_G)u({\bf p}_{1G}^*)u({\bf p}_{2G}^*)u({\bf p}_{3G}^*)$$
$$<{\bf p}_2, {\bf p}_3, {\bf V}|
\widehat B({\bf v}_G)|\psi_{RF}>= \eqno(4.5a)$$
$$=u({\bf p}_1) u({\bf p}_2) u({\bf p}_3)
\psi_G({\bf p}_2, {\bf p}_3, {\bf V})\eqno(4.5b)$$
where eqs.(3.12a,b) and (4.3) have been taken into account.
Also, equivalently, making  explicit use of eq.(3.12b), one can write
$$\psi^D_G({\bf p}_2, {\bf p}_3,{\bf V})= 
\left[{ \frac  {\epsilon({{{\bf p}^*_{2G}}}) \epsilon({{{\bf p}^*_{3G}}}) }
               {\epsilon({\bf p}_2)   \epsilon({\bf p}_3)  }  }\right]^{1/2}
\varphi^D_G({\bf p}_2, {\bf p}_3;{\bf v}_G)
(1+{\bf v}_G^2)^{-1/4}\delta({\bf V} - {\bf v}_G)   \eqno(4.5c)$$ 
with
$$\varphi^D_G({\bf p}_2, {\bf p}_3;{\bf v}_G)= 
B^D({\bf v}_G)u({\bf p}_{1G}^*)u({\bf p}_{2G}^*)u({\bf p}_{3G}^*)
\psi^{J \Sigma}({\bf p}^*_{2G}, {\bf p}^*_{3G}) \eqno(4.5d)$$
The expression $\psi^D_G({\bf p}_2, {\bf p}_3,{\bf V})$ 
of eq.(4.5c) is the boosted Dirac wave function of the model.
Also,  $\varphi^D_G({\bf p}_2, {\bf p}_3;{\bf v}_G)$
of eq.(4.5d) can be defined  
as the boosted \textit{intrisic} Dirac wave fuction.
This expression will be used in the next section 
for writing the
manifestly covariant current operators.

\noindent
Finally, we recall that, in all the previous eqs.(4.5a-d),
\textit{ the expression of the} ${\bf p}^*_{iG}$ \textit{given in} eq.(3.12c) 
\textit{must be used}.

\noindent
The previous discussion has been focussed on the boost transformation from the RF to a GF.
However, recalling the general property of eq.(4.3), 
one can immediately verify the equivalence
of our model to  SF in the case of a transformation from a GF to another GF. 
\vskip 1.0 truecm
\noindent
\textbf{5. The matrix elements of the current operators. Comparison with SF.}
\vskip 0.5 truecm
\noindent
In this section we  first examine the construction of transition matrix elements 
introducing the RIA; 
then, we critically discuss the equivalence  of our formalism with SF.
We recall that, in order to compare the
theoretical model with the experimental data,
the electromagnetic and weak 
\textit{form factors} can be easily extracted 
from the corresponding current matrix elements [1-3].

\vskip 0.5 truecm
\noindent
The main hypothesis of the RIA, as in the nonrelativistic case, 
consists in assuming  that, formally, only one constituent 
quark interacts with the external probe while the others act as \textit{spectators}.
Considering the choice of the independent momenta performed in the previous sections, 
we conveniently take 
the quark $\#1$ as the interacting one and the quarks $\#2$ and $\# 3$ as spectators.
The matrix element calculated according to this hypothesis, 
is then multiplied 
by a factor 3 to obtain the total amplitude
(when considering three identical particles).
\vskip 0.5 truecm
\noindent
In order to construct current transition matrix elements with explicit 
relativistic tensor properties,
we shall use the boosted  Dirac wave functions of eqs.(4.5a-d) 
and make, in a GF, the integrations over
${\bf p_2}$ and ${\bf p_3}$, that are the spatial variables of the spectator quarks.
According to the impulse approximation, these momenta  remain unchanged 
in the initial and final state of the scattering process.
\vskip 0.5 truecm
\noindent  
In more detail,
we  shall denote the four-momentum of the bound system, 
observed in the GF, as $P^{\mu}_G$. The index $G$ will be set to $I$ and $F$
for the initial and final state, 
respectively. The same notation will be used extensively in the following
of this section.

\noindent 
The numerical parameters $v^\mu_G$ (introduced in Sect.3) 
for boosting the wave function from the initial
or final RF to the GF, are determined by means of eq.(2.3) in the form
$$v^\mu_G=P^\mu_G/M_G \eqno(5.1a)$$ 
with 
$$M_{G}= \sqrt{P^\mu_{G} P^\nu_{G}g_{\mu \nu}} \eqno(5.1b)$$
\vskip 0.5 truecm
\noindent
As before, the independent components are the spatial ones, i.e. ${\bf v}_{G}$.
As shown in eq.(3.12b), the bound system is in an eigenstate with 
${\bf V}={\bf v}_{G}$.

\noindent
In this work we consider,
for definiteness, elastic transition amplitudes, that is with $M_G=M$, but the method 
can be generalized to the case of inelastic processes. 

\noindent
For the whole bound system, we introduce the total
(measured) four-momentum transfer  
$q^\mu$, that is 
$P^\mu_F-P^\mu_I=  q^\mu= (q^0,{\bf q}) $, and $Q^2=-q_\mu q^\mu > 0$.
\vskip 0.5 truecm
\noindent
We observe that, on the other hand, 
the four-momentum 
(denoted as $\bar q^\mu$) adquired by the interacting quark $\#1$,
that remains \textit{on shell} in the scattering process, 
can be  easily calculated from eq.(2.14) and depends on the dynamical state of the system.
Explicitly, it has the form
$$ \bar q^\mu=p^\mu_{1F}-p^\mu_{1I}=   
  {v_F^\mu} \cdot M_f({\bf p}_2,{\bf p}_3,{\bf v}_F)  
 -{v_I^\mu} \cdot M_f({\bf p}_2,{\bf p}_3,{\bf v}_I)\eqno(5.2)  $$    
At variance with the nonrelativistic impulse approximation,
$\bar q^\mu$  is not equal to the
measured momentum transfer $q^\mu$ [2,3] .   
\vskip 0.5 truecm
\noindent
According to the previous considerations, the current matrix element 
can be written in the following  general form

$$  \hat I_{FI}= 3\int d^3{\bf p}_2 d^3{\bf p}_3 d^3{\bf V} d^3{\bf V}' $$
$$ \bar\psi^D_F({\bf p}_2, {\bf p}_3,{\bf V}){\mathcal N}_F
 ~ { e}_1  ~\hat\Gamma_1 ~{\mathcal N}_I
\psi^D_I({\bf p}_2, {\bf p}_3,{\bf V}')$$
$${1 \over M} 
(1+{\bf V}^2)^{1/4}\delta({\bf V} - {\bf V}' -M{\bf q})  
(1+{\bf V}'^2)^{1/4}  \eqno(5.3a)  $$
$$=\hat J_{FI}\delta({\bf P}_F-{\bf P}_I-{\bf q}) \eqno(5.3b)$$
with
     
$$\hat J_{FI}=  3\int {d^3{\bf p}_2 \over \epsilon({\bf p}_2)} 
                 {d^3{\bf p}_3 \over \epsilon({\bf p}_3)} 
\bar\varphi^D_F({\bf p}_2, {\bf p}_3;{\bf v}_F)
[\epsilon({{{\bf p}^*_{2F}}}) \epsilon({{{\bf p}^*_{3F}}})]^{1/2} $$
$${\mathcal N}_F   ~ { e}_1  ~    \hat\Gamma_1 ~{\mathcal N}_I $$
$$[\epsilon({{{\bf p}^*_{2I}}}) \epsilon({{{\bf p}^*_{3I}}})]^{1/2} 
\varphi^D_I({\bf p}_2, {\bf p}_3;{\bf v}_I)\eqno(5.3c)$$
Let us now comment the previous expressions. 

\noindent
As anticipated, the  factor 3 that multiplies 
the matrix element, by means of the antisymmetry of the wave function, takes into account
the contributions of the quarks $\#2$ and  $\#3$, when \textit{these quarks are 
interacting} with the virtual photon field.

\noindent
The factors in the last line of eq.(5.3a) represent the matrix element, 
in the ${\bf V}$ representation, of the operator that changes 
the total momentum of the system.

\noindent
Eq.(4.5c) has been used to transform eq.(5.3a) into eqs.(5.3b,c).
The Dirac adjoint wave functions have been introduced multiplying the Hermitic conjugate
by $\gamma_1^0\otimes  \gamma_2^0\otimes  \gamma_3^0$.

\noindent
The factors ${\mathcal N}_G$ represent \textit{invariant} but, in some extent,
\textit{arbitrary} normalization functions [6] that will be briefly discussed 
in the following for the electromagnetic form factors.

\noindent
The generalized charge operator ${ e}_1$ for the interacting quark has been introduced.
The specific form of this operator, in the isospin space,
will be given in eqs.(5.5) and (5.11) for the 
electromagnetic and axial current, respectively. 

\noindent
The symbol $\hat\Gamma_1$ denotes the covariant quark interaction vertex.
It is given by 
a subset of the 16 Dirac covariant matrices for the
quark $\#1$ multiplied by spatial functions with definite Lorentz tensor properties.
We recall that one has the following Dirac matrices:
$\hat\Gamma_1= {\bf 1}_1,\gamma_1^\mu,\gamma_1^5\gamma_1^\mu$....
for scalar, vector, axial-vector.... matrix elements, respectively.

\noindent
The covariant matrix element of the model is $\hat J_{FI}$. 
To clarify the meaning of this quantity we recall that, 
for a \textit{single} (point-like) spin $1/2$ particle, it would be represented 
by the standard expression
$~\hat J_{FI}=w^+_{\Sigma_F}\bar u({\bf P}_F) { e} \hat\Gamma
u({\bf P}_I) w_{\Sigma_I}$.
\vskip 0.5 truecm
\noindent
We highlight that
our model for the current matrix elements of a composite system, represented by 
eq.(5.3c), is \textit{manifestly covariant},
according to the definition given in the introduction.
In fact, in eq.(5.3c) there appear
\textit{ covariant integrations} over 
the spectator quark momenta and
\textit{ invariant factors}.
Also, the \textit{intrinsic} RF wave functions are boosted by means
of standard Dirac boosts.
Finally, the prove of covariance is completed by using 
\textit{standard}  boost transformation properties
of the Dirac matrices. 
In particular:  

$$B^D({\bf v}) \gamma^0 B^D({\bf v})=\gamma^0 \eqno(5.4a)$$ 
$$B^D({\bf v}) \gamma^0\gamma^\mu B^D({\bf v})=
\gamma^0L^\mu_{~\nu}({\bf v}) \gamma^\nu \eqno(5.4b)$$ 
and the corresponding transformations for the other Dirac matrices.
\vskip 0.5 truecm
\noindent
The most relevant case for the study of the hadronic structure is represented by the
four-vector electromagnetic interaction. 
In this case one has the following quark charge operator in the isospin space
$${e}_1 =  {{ e}_1}^{~em} = {1\over 2}{\tau_1}^3+{1\over 6}\eqno(5.5) $$ 
The four-vector vertex can be put in the following phenomenological general form
$$\hat\Gamma_1=\Gamma_1^\mu=\gamma_1^\mu F_A - {1\over 2}\sigma_1^{\mu \nu}F_B
\cdot(K_{F}v_{\nu F} -K_{I}v_{\nu I} )\eqno(5.6)  $$
with the invariant factors
$$ F_A=F_A(M;{\bf p}_2,{\bf p}_3,{\bf v}_F,{\bf v}_I)\eqno(5.7a)$$
$$ F_B=F_B(M;{\bf p}_2,{\bf p}_3,{\bf v}_F,{\bf v}_I)\eqno(5.7b)$$
$$ K_G=K_G(M; {\bf v}_G ,{\bf p}_2,{\bf p}_3)\eqno(5.7c)$$

\noindent
In the case of a single but \textit{nonpoint-like}
(on shell) particle they represent the standard observable form factors
$$ F_A=F_1(Q^2), F_B=F_2(Q^2), K_G=1 \eqno(5.8)$$
On the other hand, when considering interacting quarks, 
$F_A$, $F_B$ and $K_G$ can give an
\textit{effective} representation of all the unknown effects
that modify the bare quark vertex.
Some of this effects can be related to violations of the RIA, others
to the strong interactions of the constituent quarks. The latter are usually
interpreted in terms of quark substructure and/or exchange of vector mesons
between the virtual photon and the quark vertex.  

\noindent
In any case, at zero momentum tranfer,  vertex charge normalization requires 
$F_A(M;{\bf p}_2,{\bf p}_3,{\bf v}_G,{\bf v}_G)=1$. The second term in eq.(5.6)
is related, at zero momentum transfer, to the quark anomalous magnetic moment. 
Furthermore, in order to represent the dependence of that term 
on the interacting quark momentum transfer. 
one can take $K_G=M_f({\bf p}_2,{\bf p}_3,{\bf v}_G)$, obtaining
 $\bar q^\mu$ as defined in eq.(5.2).

\vskip 0.5 truecm
\noindent
However, the simplest choice is to consider
the interaction of the virtual photon with point-like Dirac particles, 
that is using $F_A=1$ and $F_B=0$ in eq.(5.6). 
In our opinion a relativistic study of the nucleon form factors should calculate
\textit{first} these quantities \textit{with that choice} 
(by using the nucleon wave functions of the quark model),
\textit{ then} insert the phenomenological functions $F_A$, $F_B$ and $K_G$
to improve the reproduction of the experimental data. 

\noindent
The study of the counterterms due to  \textit{dynamical} current conservation
and the analysis of the contributions due to virtual negative energy states or to
quark-antiquark pairs can help to construct a more reliable and consistent model.

\vskip 0.5 truecm
\noindent
As for the invariant normalization factors ${\mathcal N}_G$ of eqs.(5.3a,c), 
they can be chosen considering the requirement of
\textit{total charge normalization}
for the matrix element at zero momentum transfer.
It reads $ J^0_{G G}=e^{em}_{tot}=+1,~0 $
for the proton and the neutron, respectively.
This condition  is \textit{automatically}
satisfied (considering the antisymmetry of the wave function
and the normalization of the Dirac spinors of eq.(4.1b)) by 

$$\mathcal N_G=[{m\over {\epsilon ({\bf p}_{1G}^*)}}]^{1/2} \eqno(5.9)$$

\vskip 0.5 truecm
\noindent
Within this theoretical framework,
various numerical calculations have been performed  
for the nucleon electromagnetic form factors, by using different
constituent quark nucleon wave functions. 
The results, in good agreement with the new experimental
data, show the essential r\^ole of relativity in such calculations and
the reliability of the RIA [2,7] as a starting point for the study of the
electromagnetic response of the nucleon.

\vskip 0.5 truecm
\noindent
Similar calculations have been also performed for the study of the axial 
nucleon form factor [3,4]. In this case, the quark interaction vertex is taken as
the axial-vector Dirac matrices 

$$\hat\Gamma_1= \gamma_1^5\gamma_1^\mu\eqno(5.10)$$
and the \textit{axial charge} as an isospin  raising operator,
that is

$${ e}_1 =  {{e}_1}^{ax} = {\tau_1}^+\eqno(5.11) $$ 
In this case no charge normalization condition can be found and
the form of the vertex spatial functions and of the normalizations factors 
should be carefully studied [4,6].

\vskip 0.5 truecm
\noindent
We now turn to discuss the comparison of our manifestly covariant matrix element
of eq.(5.3c) with that of the SF [2,6].
For definiteness we refer to a four-vector electromagnetic vertex 
$\hat \Gamma_1=\gamma_1^\mu$ with the normalization
factors given in eq.(5.9).

\noindent
To this aim we shall transform our expression of eq.(5.3c) into the SF.
We divide this procedure into  the following three steps.
First (i), we obtain the rotation matrices of the spectator quarks; 
second (ii),
the rotation matrices of the interacting quark; finally (iii), the momentum 
$\delta$ functions of the spectator quarks.

\vskip 0.5 truecm
\noindent
(i) We now reproduce the rotation matrices of the spectator quarks $(i=2,3)$.
For these quarks,
taking into account eqs.(3.8b) and (4.5d), the momenta of eq.(3.12c),
in eq.(5.3c) one has 
the following spinorial bilinear quantities 
$$S_{{\sigma_i}' {\sigma_i}}=w^+_{{\sigma_i}'}
u^+({\bf p}_{iF}^*)B^D_i({\bf v}_F)\gamma^0_i
B^D_i({\bf v}_I)u({\bf p}_{iI}^*)w_{\sigma_i }\eqno(5.12)$$
By means of eq.(5.4b) one can write
$$B^D_i({\bf v}_F)\gamma^0_iB^D_i({\bf v}_I)=
\gamma^0_i 
[B^D_i({\bf v}_F)]^{-1}B^D_i({\bf v}_I)=
\gamma^0_i
B^D_i(-{\bf v}_F)B^D_i({\bf v}_I)\eqno(5.13)$$
We now consider the product of the two Dirac boosts in the last equation.
We recall that the corresponding boosts on the spectator momenta are
$${\bf p}_b({\bf p}_{iI}^*; {\bf v}_I)={\bf p}_i \eqno(5.14a)$$
$${\bf p}_b({\bf p}_{iF}^*; {\bf v}_F)={\bf p}_i \eqno(5.14b)$$
The last equation can be rewritten as
$${\bf p}_b({\bf p}_i; -{\bf v}_F)={\bf p}_{iF}^* \eqno(5.14c)$$
In consequence, applying successively (\textit{composing})
the boosts of eqs.(5.14a) and (5.14c), 
one obtains the following total boost
$${\bf p}_b[ {\bf p}_b({\bf p}_{iI}^*; {\bf v}_I)
; -{\bf v}_F]={\bf p}_{iF}^* \eqno(5.14d)$$
Note that for the Dirac spinors the corresponding boost is the product
$B^D_i(-{\bf v}_F)B^D_i({\bf v}_I)$
of eq.(5.13).
We use for that product, applied to $ u({\bf p}_{iI}^*)$
the property of eq.(4.3). Then, we insert the result in eq.(5.12).
Taking
the covariant Dirac spinor normalization and the definition  of eq.(3.8) for the
Wigner rotations, one finally obtains
$$S_{{\sigma_i}' {\sigma_i}}= 
D_{{\sigma_i}' \sigma_i}^{1/2}
[ R_W({\bf p}_{iI}^*; B^{-1}({\bf v}_F) B({\bf v}_I))]$$
$$=\sum_{\lambda_i}
D^{* 1/2}_{\lambda_i \mu'_i}[R_W({\bf p}_{iI}^*,B({\bf v}_F))]
D^{  1/2}_{\lambda_i \mu_i }[R_W({\bf p}_{iF}^*,B({\bf v}_I))]
 \eqno(5.15)$$
The second equality is directly obtained without \textit{composing} the two
successive boosts.

\vskip 5.0 truecm
\noindent
(ii) As for the Wigner rotations of the interacting quark,
by means of eq.(4.3) and inserting two complete sets of spin states,
we introduce the following identity
$$   w^+_{{\sigma_1}'}
u^+({\bf p}_{1F}^*)B^D_1({\bf v}_F)\gamma^0_1 
\Gamma_1^{\mu}
B^D_1({\bf v}_I)u({\bf p}_{1I}^*)w_{\sigma_1 }$$
$$=\sum_{\lambda_1 {\lambda_1}'}
D_{{{\lambda_1}'} {\sigma_1}'}^{1/2~*}[ R_W({\bf p}_{1F}^{*},B({\bf v}_F))] $$
$$w^+ _{{\lambda_1}'}   
\bar u({\bf p}_b({\bf p}_{1F}^*; {\bf v}_F))
\Gamma_1^{\mu}
u({\bf p}_b({\bf p}_{1I}^*; {\bf v}_I))
w_{\lambda_1}$$ 
$$D_{{\lambda_1} \sigma_1}^{1/2}[ R_W({\bf p}_{1I}^{*},B({\bf v}_I))] \eqno(5.16)$$

\vskip 0.5 truecm
\noindent
(iii) Let us now consider the spatial integrations 
over the spectator momenta
of our eq.(5.3c). We introduce two $\delta$ functions in the following way:

$$\int d^3{\bf p}_2  d^3{\bf p}_3 ~~....=
  \int d^3{\bf p}_2'  d^3{\bf p}_3' 
        d^3{\bf p}_2  d^3{\bf p}_3 ~
\delta({\bf p}_2'- {\bf p}_2)\delta({\bf p}_3'- {\bf p}_3)~....
\eqno(5.17) $$
Furthermore, in eq.(5.3),  ${\bf p}_2'$ and ${\bf p}_3'$ 
are then taken as the arguments of the final state
wave function. The rest frame final momenta
are considered as functions of those momenta. The same holds for the initial state,
taking  ${\bf p}_2$ and ${\bf p}_3$ as arguments. 

\noindent
We now replace the integration variables ${\bf p}_i$ ($i=2,3$)
\textit{ and the primed ones}
with the corresponding rest frame momenta ${\bf p}_i^*$. 
For the initial state momenta, a transformation factor must be introduced according 
to the following equation

$$d^3 {\bf p}_i={\epsilon({\bf p}_i) \over 
{  \epsilon({\bf p}_i^*)}}~ d^3 {\bf p}_i^* \eqno(5.18)$$
An analogous equation holds for the final state primed momenta.
We can identify
$${\bf p}_{i}^*=   {\bf p}_{iI}^* \eqno(5.19a)$$
$${\bf p}_{i}'^*={\bf p}_{iF}^* \eqno(5.19b)$$
and use eqs.(5.14a,b), respectively, to express ${\bf p}_i$ and  ${\bf p}'_i$ as functions
of the intrinsic momenta.

\noindent 
Considering eq.(3.9), the rest frame momenta are easily replaced by the Jacobi momenta 
as integration variables.

\noindent
Collecting all the previous results,
our electromagnetic current matrix element is put in the SF, giving

$$J^\mu_{FI}= 3\int d^3{\bf p}_\rho d^3 {\bf p}_\lambda
d^3{\bf p}'_\rho d^3 {\bf p}'_\lambda 
\psi^{1/2 ~\Sigma_F~* }_{{\sigma_1}' {\sigma_2}' {\sigma_3}'}  
({\bf p}'_\rho, {\bf p}'_\lambda)~  $$
$$D_{{{\lambda_1}'} {\sigma_1}'}^{1/2~*}[ R_W({\bf p}_1'^{*},B({\bf v}_F))] 
 w^+ _{{\lambda_1}'}   \bar u({\bf p}'_1) e_1~\gamma_1^\mu
u({\bf p}_1)w_{\lambda_1} 
D_{{\lambda_1} \sigma_1}^{1/2}[ R_W({\bf p}_1^{*},B({\bf v}_I))]$$ 
$$ D_{{\sigma_2}' \sigma_2}^{1/2}[ R_W({\bf p}_2^{*}, B(-{\bf v}_F) B({\bf v}_I))]
   D_{{\sigma_3}' \sigma_3}^{1/2}[ R_W({\bf p}_3^{*}, B(-{\bf v}_F) B({\bf v}_I))]$$
$$\delta({\bf p}_2' -{\bf p}_2 )
  \delta({\bf p}_3' -{\bf p}_3 )\cdot
m\epsilon({\bf p}_2)\epsilon({\bf p}_3)$$
$$[\epsilon({\bf p}_1'^*)\epsilon({\bf p}_2'^*)\epsilon({\bf p}_3'^*)
   \epsilon({\bf p}_1^*)\epsilon({\bf p}_2^*)\epsilon({\bf p}_3^*)]^{-1/2}$$
$$  \psi^{1/2 ~\Sigma_I }_{\sigma_1 \sigma_2 \sigma_3} 
({\bf p}_\rho, {\bf p}_\lambda) 
\eqno(5.20)$$

\noindent
where a sum over the repeated indices is understood.

\noindent
We note that the previous expression, that has been shown to be equal to eq.(5.3c) 
for the electromagnetic interaction, is 
\textit{cohincident} with eqs.(2),(3) and (10)
of ref.[6]. 
Apart from a (probably not relevant) normalization factor, our expression is 
also equivalent to the result of ref.[2].
\vskip 0.5 truecm
\noindent
After verifying the equivalence of our covariant matrix element with SF, 
we conclude observing that our expression of eq.(5.3c) presents the following 
advantages with respect to SF.

\noindent
(i) As discussed above, it is \textit{manifestly covariant}.

\noindent
(ii) It is more compact, in the sense that it contains 
only \textit{two} three-dimensional integrations over the spectator momenta
with respect to \textit{four} integrations of the SF.

\noindent
(iii) Well known Dirac spinors and Dirac boost matrices are used instead of
rotation matrices of Wigner rotations.

\vskip 0.5 truecm
\noindent
These features  allow for studying the possibility of deriving an expression for
a \textit{ dynamically  conserved current} by means of a suitable procedure of
minimal coupling substitution. The results of this investigation will be presented in
subsequent works.

\vskip 1.0 truecm
\noindent
\textbf{Acknowledgments}
\vskip 0.5 truecm
\noindent
I thank Prof. M.M. Giannini, Dr E. Santopinto and Dr. A. Vassallo 
of INFN Sez. di Genova - Italy, for critical discussions and, in particular, 
for having suggested 
to compare the manifestly covariant formalism with the SF of the PF RHD.


\end{document}